\documentclass[a4paper,11pt]{article}
\usepackage{pos}
\usepackage{lmodern}
\usepackage[
  natbib=true,
  backend=bibtex,
  sorting=none
  ]{biblatex}
\usepackage{layouts}
\usepackage{bookmark}

\title{On the Transients Handler for the Cherenkov Telescope Array Observatory}
\author*[a]{Tiffany Collins}
\author[a]{Kathrin Egberts}
\author[b]{Constantin Steppa}
\author[c]{Igor Oya}
\author[b]{Dominik Neise}
\onbehalf{for the CTAO Consortium}

\affiliation[a]{Instit\"ut f\"ur Physik und Astronomie, Universit\"at, Potsdam, Karl Liebknecht-Stra{\ss}e 24-25, Potsdam, Germany}

\affiliation[b]{CTAO gGmbH, Saupfercheckweg 1, 69117 Heidelberg, Germany}

\affiliation[c]{Centro de Investigaciones Energ\'eticas, Medioambientales y Tecnol\'ogicas (CIEMAT), Av. Complutense 40, Madrid, Spain}

\emailAdd{tiffany.collins@uni-potsdam.de}

\abstract{The Cherenkov Telescope Array Observatory (CTAO) is a next-generation gamma-ray observatory in both the Southern (Paranal, Chile) and Northern Hemisphere (La Palma, Spain) and will consist of up to 100 imaging atmospheric Cherenkov telescopes. With sensitivity far exceeding current facilities, CTAO will provide detailed measurements of gamma rays from GeV up to a few 100s of TeV. CTAO has a nominal field of view of ~10º and will rely on external alerts from observatories such as IceCube (neutrinos), Fermi-LAT (GeV gamma rays), LIGO (gravitational waves (GWs)) or the upcoming Vera C. Rubin Observatory (optical) to trigger observations of targets of opportunity (ToOs).
The Transients Handler (TH) is a subsystem of the Array Control and Data Acquisition that will provide the means of handling alerts and schedule follow-up observations. The TH is responsible for (i) filtering of thousands of internal and external alerts per night, (ii) processing of alerts according to the corresponding observation proposal to determine the optimal observation schedule with minimum interruptions to observations and (iii) broadcasting of events detected by CTAO to external facilities. In this work, we will discuss the architecture of the Transients Handler, its latest improvements and future updates.}

\ConferenceLogo{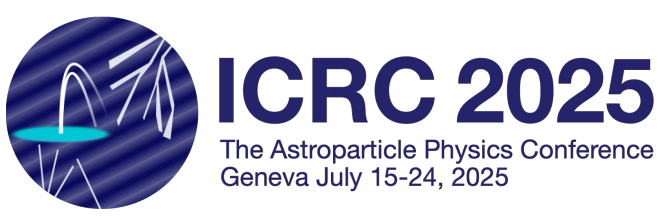}

\FullConference{39th International Cosmic Ray Conference (ICRC2025)\\
 15-24 July 2025\\
Geneva, Switzerland\\}
\begin{document}
\maketitle

\section{Introduction}
The Cherenkov Telescope Array Observatory (CTAO) \cite{CTAbook} with its three different classes of telescopes will be able to observe the gamma-ray Universe in more detail than ever before. A key science project of CTAO is to follow up on targets of opportunity (ToO) to understand the underlying science behind transient events. Transient events are ToOs that flare up across the electromagnetic spectrum and other messangers (neutrinos, gravitational waves) over a time scale of milliseconds to weeks. These transient events are associated with some of the most extreme environments in the Universe. In order to observe transients, CTAO will rely on the Transients Handler to automatically process incoming alerts from external instruments within four seconds and determine the optimum observation strategy. The Transients Handler \cite{10.1117/12.2629372,10.1117/12.3017539} is designed to determine the optimal observation strategies for detecting transients while adapting to the amount of incoming alerts and the priority of the alerts based on science configurations.

\section{The Transients Handler}

\begin{figure}[t!]
    \centering
    \includegraphics[width=0.8\textwidth]{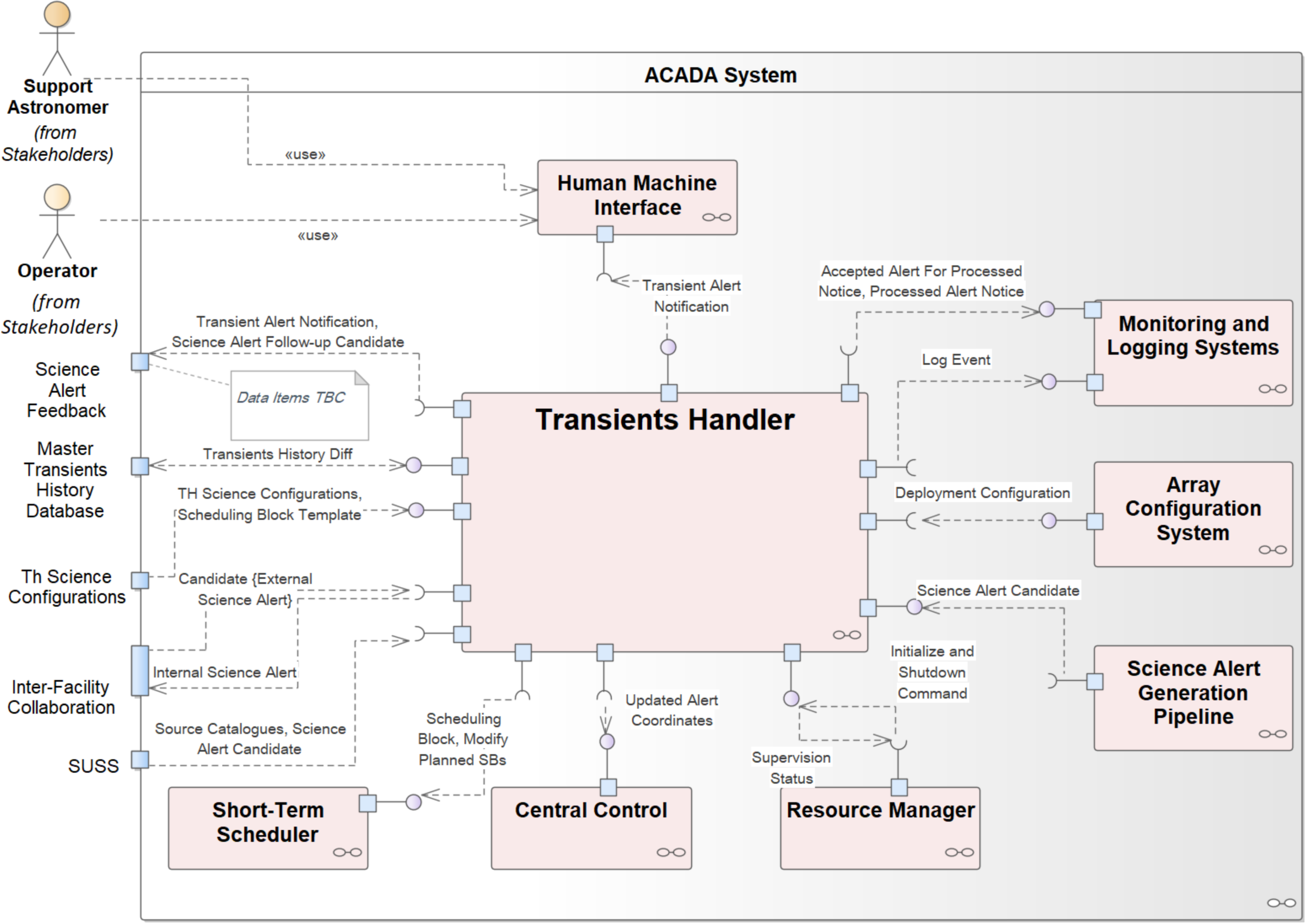}
    \caption{Context diagram for the Transients Handler of the Array Control and Data Acquisition System (ACADA).}
    \label{fig:ACADA}
\end{figure}

The Transients Handler is a subsystem of the Array Control and Data Acquisition System (ACADA) \cite{2024SPIE13101E..1DO}, which is the central software that controls, supervises and handles the collection of data of the telescopes at both CTAO sites. ACADA consists of multiple subsystems that is written in different languages (Python, Javascript, Java and C++) and is based on the ALMA Control Software (ACS) framework \cite{2002SPIE.4848...43C}. The Transients Handler receives alerts from external instruments and alerts generated by the Science Alert Generation (SAG) pipeline \cite{2022icrc.confE.937B}. If the alert is accepted and a scheduling block is created, the Transients Handler sends the alert to the Short-Term Scheduler (STS) which then schedules observations automatically (see \autoref{fig:ACADA}).

\begin{figure}[t!]
    \centering
    \includegraphics[width=\textwidth]{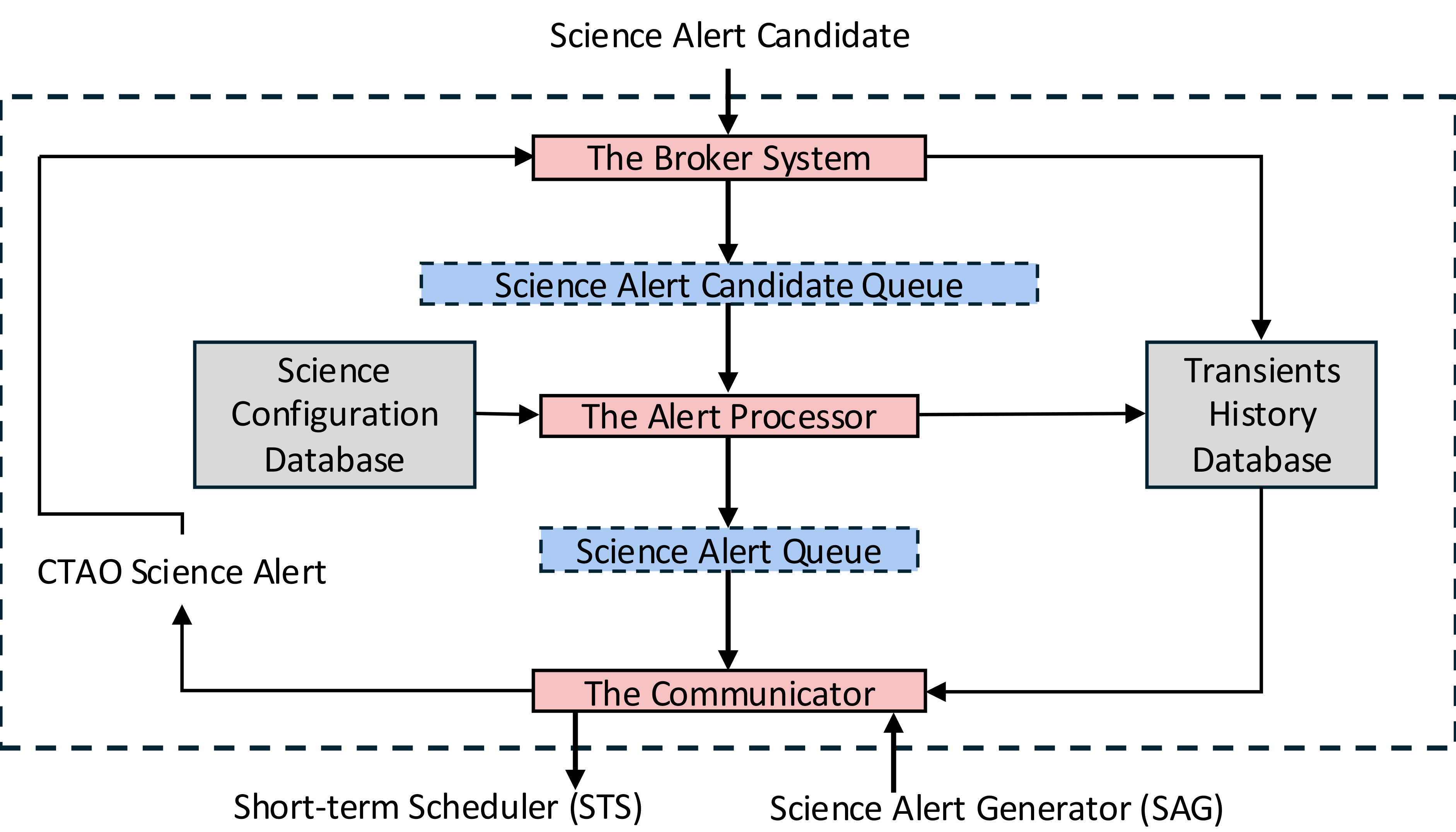}
    \caption{Design of the Transients Handler.}
    \label{fig:TH}
\end{figure}

The Transients Handler is written using Python 3.12.0 and consists of three main components: the Broker Manager, the Alert Processor and the Communicator (see \autoref{fig:TH}). It makes use of two databases implemented with MongoDB \cite{MongoDB}, the Science Configuration database and the Transients History database. The Science Configuration database contains the science configurations that correspond to different transient events and contains instructions on how to filter and process the alert. The Transients History database contains the history of all incoming alerts, whether the alert was accepted by the TH, the content of the original alert and the processing results including the proposed scheduling block. The components of the Transients Handler are linked to each other by two persistent queues: the Science Alert Candidate (SAC) queue and the Science Alert (SA) Queue. These queues allow alerts with high priority (as defined by the science configuration) to be processed and communicated before those of lower priority.

\subsection{The Broker Manager}

The Broker Manager of the Transients handler is responsible for maintaining the multiple brokers needed for CTAO to receive and send alerts. The Broker Manager will consist of three different types of brokers: internal brokers, external brokers and an alert broadcaster. Internal brokers interface with the SAG pipeline and SUSS to receive internal alerts generated by CTAO. If an internal alert is detected, the alert broadcaster broker will alert external instruments about the ToO. The Broker Manager will connect to external instruments and the General Coordinates Network (GCN) \cite{2024AAS...24335919S} to receive external alerts. These alerts are in VOEvent2.0 format which is the current international (IVOA) standard for reporting transients astrophysical phenomena \cite{2011ivoa.spec.0711S}. The Broker Manager provides basic filtering of the alerts (e.g. based on keywords) and creates an entry in the Transients History database containing the original content of the alert. The alert is considered a Science Alert Candidate and passed onto the SAC queue. The lastest release of the Transients Handler makes use of the Comet broker \cite{SWINBANK201412} to connect to GCN and to listen to alerts from Fermi-GBM \cite{2009ApJ...702..791M} and Swift-BAT \cite{Krimm_2013}.

\subsection{The Alert Processor}

The Alert Processor pulls alerts from the SAC queue and processes the alert to determine the optimum observation strategy. The stages of processing are as following: (i) The Alert Processor reads the content of the original alert from the transients history database and matches the alert to the science configurations. If an alert matches to multiple configurations, the alert will be separated into individual alert-configuration pairs. (ii) For each alert-configuration pair, the alert is filtered and processed according to the instructions given in the science configuration. This can include instructions to gather additional information required to process the alert and how to calculate the observation window. These results are then stored in the transients history database. (iii) Finally, the Alert Processor merges the results from each alert-configuration pair. If the alert has been successfully filtered and processed according to current conditions (e.g. weather, moonlight), the alert is now considered a Science Alert and passed on the SA queue. Multiple instances of the Alert Processor can run simultaneously on different threads/processes to handle the different influx of alerts from the Broker Manager.

\subsection{The Communicator}

The Communicator pulls the alert from the SA queue, creates the scheduling block based on the processing results and submits the alert to STS. It is also responsible for notifying HMI about incoming alerts and monitoring the status of the Broker Manager and Alert Processor. In a future release, the Transients Handler will provide an interface to publish alerts through the Broker Manager. 

\section{Status and Outlook}

The first version of the Transients Handler was released with ACADA version 1.0 in July 2023 and successfully tested with the first Large-sized Telescope (LST1) of CTAO in October 2023 \cite{2024SPIE13101E..0HL}. A fake alert was submitted to ACADA, which successfully scheduled and executed a ToO observation with LST. ACADA version 1.5 was released in December 2024 and implemented several of the features described above, including the addition of the SAC and SA queues.

The next release of the Transients Handler will be released with ACADA 2.0 at the end of 2025. The Transients Handler will transition from the Comet broker to the Kafka broker maintained by GCN \cite{2022GCN.32419....1B} and will include neutrinos and gravitational waves as science use cases. Consequently, the functionalities of the Alert Processor will be expanded to process these cases. Gravitational Wave alerts currently have large location uncertainties, hence the Alert Processor will make use of the python package Tilepy \cite{Seglar-Arroyo_2024} to calculate scheduling blocks. Release 2 of ACADA will consider whether STS is in manual mode, where the support astronomer, through the Human Machine Interface (HMI), can create and execute scheduling blocks. During manual mode, the Transients Handler will notify HMI of incoming alerts that have passed criteria. If the astronomer accepts this alert, the Transients Handler will then send the scheduling block to STS. The Communicator must be adapted accordingly to consider the mode of STS before the sending the alert.

\section{Conclusion}

The Transients Handler is a subsystem of ACADA that is responsible for the processing of internal and external alerts within four seconds of receival. It calculates whether a ToO can be followed up by CTAO and determines the optimal observation strategy based on current conditions with minimal interruptions to current observations. The Transients Handler provides CTAO with the means on following up on transients events in order to understand the underlying science behind some of the most extreme events in the Universe.

\section{Acknowledgements}

This work has been funded by the German Ministry for Education and Research (BMBF).

\printbibliography

\end{document}